\documentclass{mem}
\usepackage{natbib}\usepackage{txfonts}\usepackage{balance}
\usepackage{graphicx}
\usepackage[a4paper]{hyperref}
\idline{75}{282}
\begin{document}
\def\teff{$T\rm_{eff }$}
\def\kms{$\mathrm {km s}^{-1}$}

\title{
Activity induced by Gravitational Interaction in Galaxy Pairs }

   \subtitle{Mixed (E+S) morphology pairs}

\author{
D. \,Dultzin,\inst{1} J.J. \,Gonzalez,\inst{1} Y.
\,Krongold,\inst{1} H. \, Hern\'andez-Toledo,\inst{1} E.M.
\,Huerta,\inst{1} I. \,Cruz-Gonz\'alez,\inst{1} L.
\,Olgu\'{\i}n,\inst{1} P. \,Marziani,\inst{2} F.
\,Hern\'andez-Ibarra,\inst{1}}

  \offprints{D. \,Dultzin}

\institute{Instituto de Astronom\'{\i}a, UNAM, Apdo. Postal 70-264,
C.P. 04510, M\'exico, D.F. \email{deborah@astroscu.unam.mx} \and
INAF, Osservatorio di Padova, Italia
\email{paola.marziani@oapd.inaf.it} }

\authorrunning{Dultzin }

\titlerunning{environmentaly induced activity}

\abstract{ A systematic study of the nuclear emission of a sample of
97 spirals in isolated galaxy pairs with mixed morphology (E+S)
shows that: 1) AGN activity is found in 40\% of the spiral galaxies
in these pairs, 2) Only one out of the 39 AGN found has type 1
(Broad line Component) activity, and 3) AGN tend to have closer
companions than star forming galaxies. These results are at odds
with a simple Unified Model for Seyferts, where only
obscuration/orientation effects are of relevance, and neatly support
an evolutionary scenario where interactions trigger nuclear
activity, and obscuration/orientation effects may be complementary
in a certain evolutionary phase.

  }
\maketitle{}
\section{Introduction}
\label{sec:intro}

One of the outstanding problems in the understanding of Active
Galactic Nuclei (AGN) phenomenon is the role of the cicumgalactic
environemnt in the triggering of the central engine. In an attempt
to elusidate this question, in the past 10 years several efforts
have focused in the study of the environment of AGNs, from a few
kiloparsecs around the galactic nucleus to some hundreds of
kiloparsecs around the host galaxy. Most of the investigations have
dealt with samples of Seyfert galaxies, because these are the
closest clearly non-thermal dominated active nuclei. LINERs are easy
to observe, however, the nature of the dominating emission mechanism
is not well established yet (Krongold et al 2003;
Gonz{\'a}lez-Mart{\'{\i}}n et al. 2006).

It has been suggested that Seyfert 2 galaxies are in interaction
with the same frequency than star-forming galaxies (Krongold et al.
2002), while Seyfert 1 galaxies are in interaction less frequently,
comparably as often as non-active galaxies (Dultzin-Hacyan et al.
1999, Krongold et al. 2001). The most recent studies confirm these
findings considering physical companions, i. e. not only from
companions selected frrom statitiscal considerations but from actual
measurements of radial velocities for the neighboring galaxies
(Koulouridis et al., 2006a 2006b).

In all the above analyses the environment of well defined samples of
active vs. non-active galaxies were compared. In the present paper
we adopt a complementary approach. We study the incidence of nuclear
activity in a well defined sample of interacting galaxies. We focus
on the sample of isolated mixed-morphology galaxy pairs studied by
Hern\'andez-Toledo et al. (1999, 2001), obtained from the Catalog of
Isolated Pairs in the Northern Hemisphere (KPG; Karachentsev 1972).

These pairs are a unique laboratory to study the effect of tidal
forces in triggering nuclear activity because they are relatively
simple systems where a gas rich galaxy interacts with a gas poor
one. In such systems a clean interpretation of the origin and
evolution of the gaseous component is possible.

In this paper we present high resolution, long-slit, spectroscopy
for the spiral component of 97 mixed morphology pairs from the {\it
  Catalog of Isolated Pairs} (CPG; Karachentsev 1972). For details on
the properties of the mixed pair sample see Hern\'andez-Toledo et
al. (1999, 2001), and Gonzalez et al. (2008, herafter G08).

\section{Observations and Data Reduction}
\label{sec:obs}

The spectroscopic observations were carried out at the Observatorio
Astronomico Nacional \footnote{The Observatorio Astron\'omico
Nacional is operated by the Instituto de Astronom\'{\i}a of the
Universidad Nacional Aut\'onoma de M\'exico at Sierra de San Pedro
M\'artir, Baja California, M\'exico.} in San Pedro M\'artir, Baja
California, Mexico. The spectra covered the wavelength interval
$\lambda$5700-7750{\AA} at a 4.6{\AA} resolution. The data reduction
was carried out using {\sc xvista} \footnote{XVISTA is distributed
by Jon Holtzman at Nuevo M\'exico State University:
http://astro.\-nmsu.\-edu/~holtz/vista} data reduction package using
standard procedures. Details about the observations and data
reduction procedures can be found in G08.

\section{Actvity Diagnostic Criteria}

Since we lack the O III/H$_\beta$ ratio (due to the wavelength
coverage) we are unable to distinguish between Seyferts and LINERs.
Thus, we labelled both as AGN. On the other hand,  we are also
unable to distinguish between Starburst and normal star-forming
galaxies. Therefore this group is considered as non-active and is
hereafter referred to as ``normal galaxies''

To classify the spiral galaxies into AGN and normal galaxies we have
definided an ``Activity Type Index'' (hereafter ATI) based on a
combination of the diagnostic line ratios found in the literature
(see G08 for details). The ATI takes into account a linear
combination of the SII/H$_\alpha$, NII/H$_\alpha$, and
OI/H$_\alpha$. Each of these line ratios is weigthed by its S/N. If
the significance of a given line ratio is less than 3$\sigma$ the
weight was set to zero. Otherwise, the weight was calculated using
the square of the S/N. See G08 for full details.

\begin{figure}[!t]
  \includegraphics[width=\columnwidth]{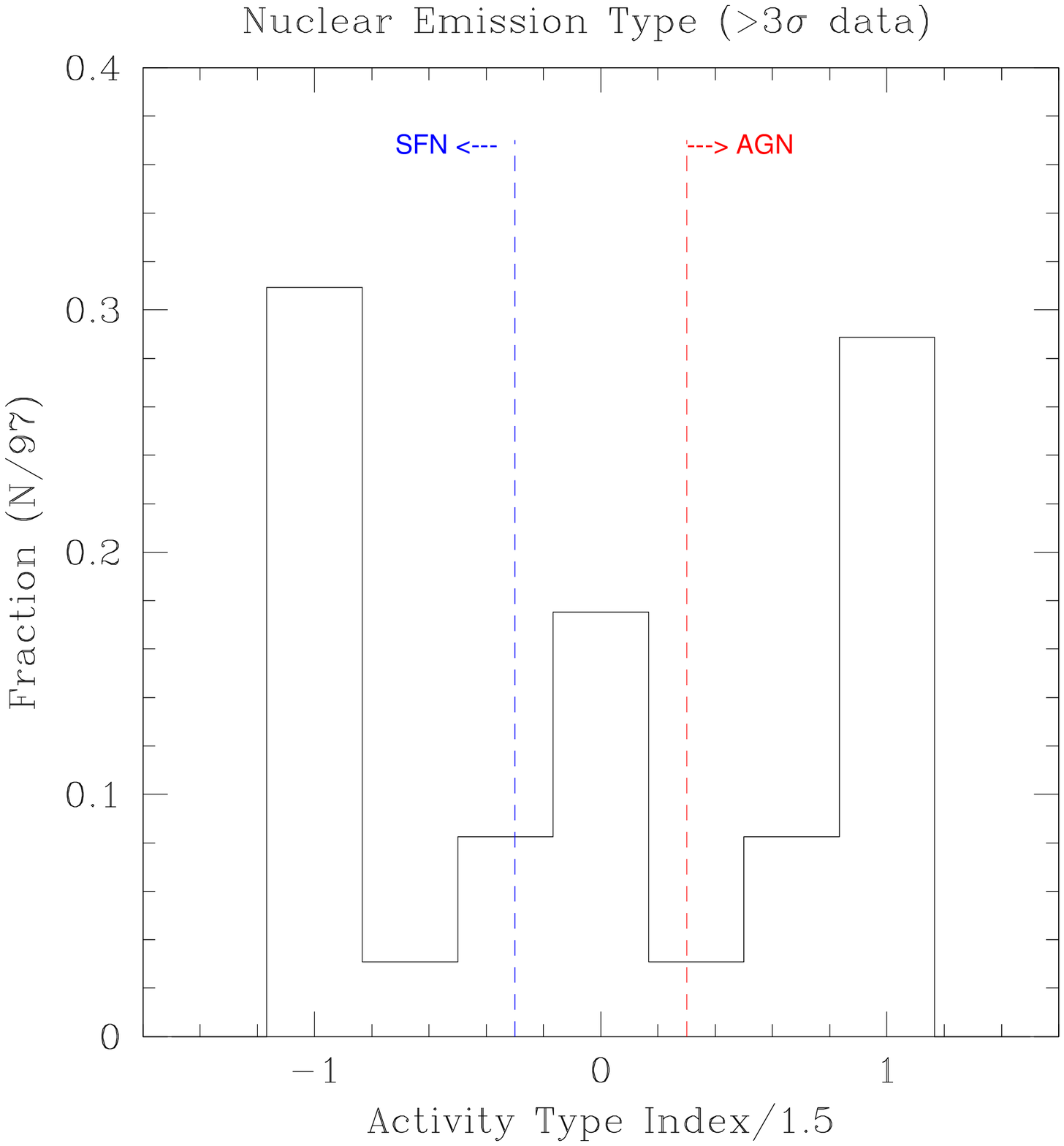}
  \caption{l}
  \label{Figure1}
\end{figure}

In Figure \ref{Figure1} the distribution of the ATI for the 97
nuclei is presented. Objects with ATI below -0.3 were considered
``normal galaxies''. Objects with ATI above 0.3 were considered
clear cut AGNs. In the range between these thresholds, it is not
possible to separate among pure AGN, composite AGN+Starburst or pure
Starburst systems. Therefore objects in this range are referred to
as ``unclassified systems''.

\section{Results}
\label{sec:results}

Figure 1 shows that 40\% of the spirals in mixed pairs show AGN
activity. This percentage is clearly higher than than the the one
found in field (non-interacting) galaxies, which is usually
estimated around 10\%.

We further check the fraction of type 1 vs. type 2 AGN in our
sample. An outstanding result is that only 1 out of the 39 objects
classified as AGN in our analysis shows the presence of a broad
H$\alpha$ component, and thus is a Type 1 object.

Finally, we examined the  relation between the ATI and the
separation of the galaxies in the pairs (normalized by the spiral
galaxy diameter measured down to the 25th isophote). The result is
shown in Fig. 2. We can see a clear cut difference between the
distribution of the separations between ``normal'' galaxies (SFN
stands for galaxies with Star Forming Nuclei), and galaxies with
active nuclei (AGN). The difference is in the sense that the Spiral
galaxies that harbour an active nucleus tend to be closer to their
elliptical neighbour. (a KS test gives a 99\% confidence level) .

\begin{figure}[!t]
\includegraphics[width=\columnwidth]{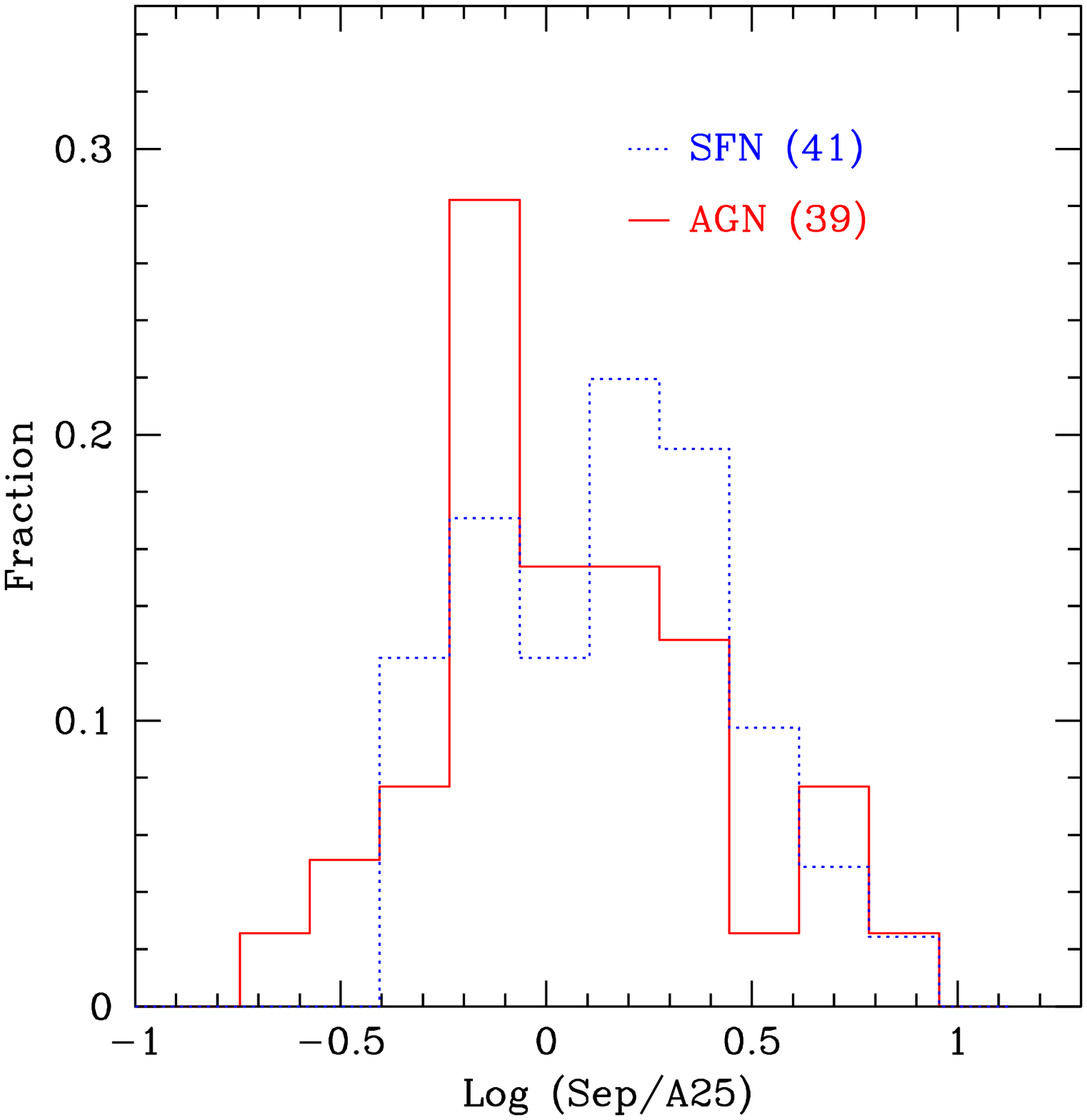}
\caption{.} \label{Figure2}
\end{figure}

\section{Discussion}
\label{sec:discussion}

The most striking result in our analysis is that only  1 out of the
39 AGN  can be classified as type 1. Even if in the present analysis
we cannot separate LINERs from Seyferts, this result is very much at
odds with the expectation of the so-called Unified Model (UM). A
frequency of 2.6\% for the presence of type 1 activity is too low to
be explained with an obscuration\/orientation effect alone. Ho et
al. (1997) found that $\sim$20\% of their sample of nearby
low-luminosity AGN (including LINERs and Seyferts) observed with HST
presented broad lines (the expectation of the UM is that $\sim$60\%
of Seyferts should be type 1).

Our results also show a clear connection between interaction and
nuclear activity, in particular of type 2 activity. This fact
follows naturally from the evolutionary model developed by Krongold
et al. (2002, 2003). In this model, the onset of both nuclear
activity and enhanced nuclear star formation is due to the tidal
forces that induce the infall of large amounts of gas into the
nucleus. These tidal effects are produced by an interaction with a
nearby (similar mass) companion.

In the first stages of the interaction, a Starburst should dominate
the emission. As more material falls into the innermost regions, the
onset of non-thermal activity begins. At this stage, the non-thermal
power is low, and there is full obscuration of the nuclear region
(including the broad line clouds), therefore only a type 2 AGN is
observed. During this stage, the model predicts a high probability
for the detection of a nearby companion. As even more material
falls, the accretion rate is increased, and partial obscuration is
expected (as the initial spherical distribution of obscuring dust
should flatten to form a torus), so both types of AGN can be
observed. In the final stage, the dust is sweeped away by both the
winds of massive evolved stars and the AGN itself. In this stage a
naked type 1 AGN neatly emerges. A timescale of 10$^9$ yr is
required to erase all evidence of possible past interactions (either
through the completion of a merging event between two galaxies of
similar mass, or through the dissolution of an unbounded interacting
companion). This naturally explains why type 2 AGN are often found
in interacting systems, while type 1 AGN are rarely found with
companions. We notice that this evolutionary scenario does not
exclude the effect of a possible obscuring/inclination combined
effect, such as the one postulated in the UM. Our scenario includes
the UM, but only at a given evolutionary stage.

The above evolutionary scenario explains in a straight forward way
the results found in this work for the spiral component of the mixed
morphology pairs. This implies that interactions play a key role in
triggering nuclear activity.

\end{document}